\documentclass{article}
\newcommand{\hred}[1]{\textcolor{red}{#1}}
\newcommand{\hgrn}[1]{\textcolor{green!50!black}{#1}}

\usepackage[final]{neurips_2026}

\usepackage[utf8]{inputenc}
\usepackage[T1]{fontenc}
\usepackage{hyperref}
\usepackage{url}
\usepackage{booktabs}
\usepackage{amsfonts}
\usepackage{amsmath}
\usepackage{amssymb}
\usepackage{nicefrac}
\PassOptionsToPackage{expansion=false}{microtype}
\usepackage{microtype}
\usepackage{xcolor}
\usepackage{graphicx}
\usepackage{multirow}
\usepackage{array}
\usepackage{caption}
\usepackage{subcaption}

\title{Single-Configuration Attack Success Rate Is Not Enough: Jailbreak Evaluations Should Report Distributional Attack Success}

\author{%
\begin{tabular}{ccc}
Carsten Maple & Abhishek Kumar & Riya Tapwal \\
University of Warwick & The Alan Turing Institute & IIT Mandi \\
\texttt{cm@warwick.ac.uk} &
\texttt{akumar@turing.ac.uk} &
\texttt{riya@iitmandi.ac.in}
\end{tabular}
}

\begin{document}

\maketitle

\begin{abstract}

Many jailbreak attack research papers report attack success rates for a limited number of parameter settings, even though there are many combinations of parameter settings that could be used. Further, when new jailbreak papers are released, they often benchmark results against single configurations of existing attacks. This position paper argues such practices are fundamentally insufficient for characterising the threat posed by parameterised jailbreak attacks, and comparing attacks. Most jailbreak attacks expose multiple internal parameters, system prompt templates, conversation rounds, cipher dispersion, teaching shots, and ASR varies substantially across these parameters. Reporting only the best-case configuration discards two pieces of information that defenders genuinely need: how typical that performance is across the variant space, and how much of the attack surface is missed by selecting a single variant. We propose two new measures for jailbreak attacks: the Variant Sensitivity Measure (VSM) and Union Coverage (UC). VSM quantifies how far the best reported ASR deviates from the mean ASR across the tested variant space, UC is the total fraction of prompts resulting in unsafe responses across all tested configurations.  We empirically demonstrate the importance of these measures using two attack families across three open-source target models. For PAIR attacks with three system prompt templates, the best single template (Authority Endorsement) achieves 69\% ASR on Mistral-7B (VSM = 0.425, UC = 88\%) and 75\% ASR on Qwen3-0.6B (VSM = 0.312, UC = 93\%). For bijection learning on Mistral-7B, the best variant achieves 81\% ASR while the union across all 36 tested variants covers 100\% of HarmBench-100 prompts, with full-space VSM of 0.803 confirming the headline overstates typical performance by a factor of five. We argue that distributional reporting, publishing VSM alongside ASR and enumerating variant coverage as fully as compute allows, should become the new minimum standard for parameterised jailbreak evaluation.

\end{abstract}


\section{Introduction}
\label{sec:introduction}


The standard practice in the LLM jailbreak literature is to report Attack Success Rate (ASR) for a limited set of attack configurations. New attack papers often present headline statements such as \textit{``our method achieves 88\% ASR on Vicuna''} or \textit{``86.3\% ASR on Claude 3.5 Sonnet,''} and subsequent work commonly treats these numbers as representative baselines when comparing against prior attacks. This convention is convenient, but it rests on a hidden assumption: that the reported configuration adequately represents the attack family as a whole. For parameterised jailbreak attacks, this assumption is often fragile, because the measured ASR may depend strongly on which system prompt template, conversation depth, encoding choice or teaching-shot count was selected.

\textbf{This position paper argues that single-configuration ASR reporting is not wrong, but incomplete.} A single number captures one configuration, whereas modern jailbreak attacks are parameterised families. PAIR attack \citep{chao2025jailbreaking} requires choosing one of three system prompt templates, the number of conversation rounds, the number of parallel streams, and the attacker LLM. Bijection learning \citep{huang2024endless} requires choosing a mapping type, dispersion level, number of teaching shots, and sampling budget. GCG \citep{zou2023universal} depends on suffix length, optimisation steps, and random seed. AutoDAN \citep{liu2023autodan} depends on population size, generations, and mutation strategy. For such attacks, a reported headline ASR may therefore reflect a particular choice of configuration, rather than the attack's behaviour across the broader parameter space.


This creates a direct problem: a single reported ASR can underestimate the total attack surface. Different variants of the same attack may jailbreak different prompts, so the prompts vulnerable to at least one variant can be substantially larger than the prompts vulnerable to the best single variant alone. A defender who focuses only on the best reported configuration may therefore miss prompts that can still be jailbroken by other variants of the same attack family.

\begin{figure}[htbp]
    \centering
    \includegraphics[width=\textwidth]{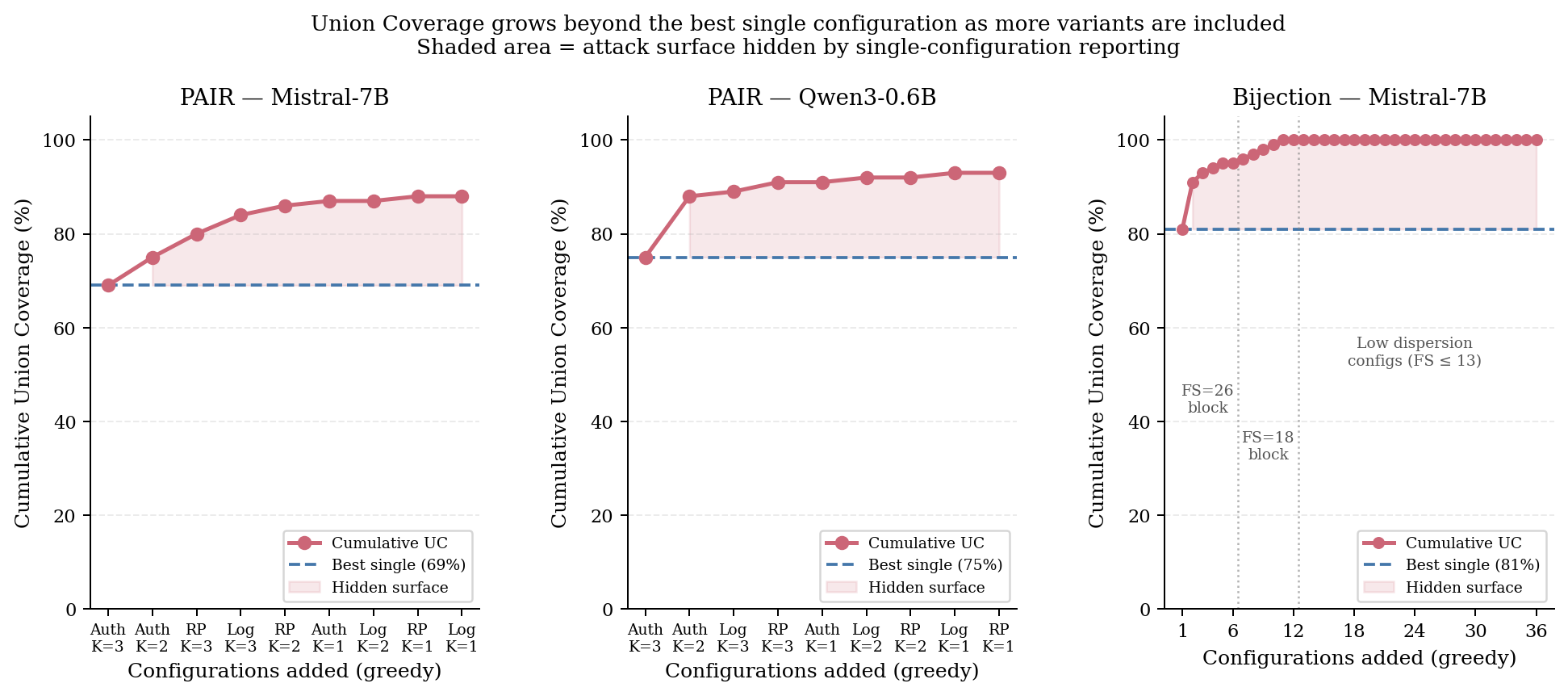}
    \caption{Cumulative Union Coverage across tested configurations. Each panel starts from the best single configuration and greedily adds configurations that contribute new successful jailbreaks. The dashed line marks the best single-configuration ASR, while the shaded region shows the additional attack surface revealed by considering multiple variants.}
    \label{fig:introoo}
\end{figure}

We present empirical evidence that this gap is large in practice. Figure~\ref{fig:introoo} illustrates the Union--Best gap using results from our PAIR and bijection experiments conducted on 100 prompts from the HarmBench dataset \cite{mazeika2024harmbench}. In each panel, the dashed horizontal line shows the best single-configuration ASR, while the solid curve shows how Union Coverage grows as additional configurations are greedily added. The shaded region marks prompts missed by single-configuration reporting: prompts not captured by the best reported configuration but jailbroken by other tested variants. Across two attack families and three open-source target models, Union Coverage exceeds the best single configuration by 18 to 33 percentage points. For PAIR, adding templates and depths beyond the best single configuration increases coverage from 69\% to 88\% on Mistral-7B~\cite{mistral7b} and from 75\% to 93\% on Qwen3-0.6B~\cite{yang2025qwen3}. For bijection learning on Mistral-7B, Union Coverage reaches 100\% across all 36 tested configurations, while the best single variant reaches only 81\%. These results show that single-configuration ASR can substantially understate the reachable attack surface.

We therefore propose two metrics, Variant Sensitivity Measure (VSM) and Union Coverage (UC), to complement ASR for parameterised jailbreak attacks. VSM measures how far the best reported ASR deviates from the mean ASR across the tested variant space, while UC measures the fraction of prompts that elicit unsafe responses when taking the union over all tested configurations. Together, these metrics indicate whether a headline ASR is representative of the evaluated variant space and how much attack surface is missed by selecting a single configuration.

\textbf{Contributions}. This paper makes the following contributions.

(1) We identify a reporting failure in current jailbreak evaluation: for parameterised attack families, selected ASR values, whether presented as a single headline number or as limited ablations, do not reveal whether the reported configurations are representative, fragile, or incomplete in terms of prompt coverage.

(2) We argue for a distributional reporting standard in which papers report the evaluated variant space, per-variant ASR, Best ASR, Mean ASR under an explicit variant distribution, Variant Sensitivity Measure (VSM), and Union Coverage (UC).

(3) We provide two empirical case studies, PAIR and bijection learning, showing that variant choice can substantially change both measured ASR and prompt-level coverage, and that different variants can expose different vulnerable prompts.

(4) We explain how this reporting standard would improve interpretation for defenders, attack researchers, and benchmark designers by distinguishing peak single-configuration performance from mean performance and cumulative prompt-level exposure.

Our claim is not that existing ASR numbers are wrong, but that selected ASR numbers are incomplete unless accompanied by the distributional and prompt-overlap information needed to interpret them.

\section{Background and Related Work}
\label{sec:related}

\subsection{Jailbreak Attacks on Large Language Models}

Adversarial attacks on aligned LLMs aim to elicit content the model has been trained to refuse. Existing attack families fall broadly into three categories: iterative LLM-driven attacks that use a separate model to refine prompts \citep{chao2025jailbreaking, mehrotra2024tree}; encoding-based attacks that obscure harmful intent through ciphers or transformations \citep{huang2024endless, ren2024codeattack}; and gradient-based optimisation attacks that directly compute adversarial suffixes \citep{zou2023universal, liu2023autodan}. Each of these methods exposes parameters, system prompt templates, search budget, dispersion, suffix length, attacker model, that materially affect attack success. 

\subsection{Existing Practice on Variant Reporting}


Some jailbreak papers report variation across selected attack parameters. PAIR \citep{chao2025jailbreaking} includes ablations on attack budget ($N \times K$) and attacker model choice, while bijection learning \citep{huang2024endless} reports ASR curves across dispersion values. CodeAttack \citep{ren2024codeattack} is more transparent than most: it reports results across three input encodings (string, queue, stack) and shows that programming language choice can substantially affect ASR, with Claude-2 increasing from 24\% ASR under Python to 74\% under Go. This is closer to the distributional reporting we advocate. However, even CodeAttack reports a single best-case stack number in its abstract and does not formalise the gap between best, mean, and union coverage as a metric. These examples show that parameter sensitivity is known, but current reporting treats such results mainly as ablations for parameter selection rather than as a standard part of attack evaluation.

\subsection{Other Critiques of Jailbreak Evaluation}

Adjacent work has raised related concerns about jailbreak evaluation. \textit{Bag of Tricks} \citep{xu2024bag} shows that external configuration choices, including model size, system prompt, chat template, attacker LLM, and attack budget, materially affect ASR. StrongREJECT \citep{souly2024strongreject} argues that judge unreliability can inflate published ASR. HarmBench \citep{mazeika2024harmbench} standardises prompts and judges to improve cross-paper comparability, but does not address within-attack variant selection. Our work is complementary: \textit{Bag of Tricks} studies between-attack experimental settings, whereas we study variation within a single parameterised attack family while holding the external setup fixed.

Our framing is also related to critiques of single-metric reporting in adversarial vision. \citet{curl2026beyond} show that ASR alone is insufficient for characterising adversarial threat in medical image classifiers, because it can be decoupled from other important properties such as perceptual quality and perturbation magnitude. We bring the same structural concern to LLM jailbreak evaluation: a single ASR value is not wrong, but it is incomplete unless accompanied by distributional information about the tested variant space.



\section{Proposed Metrics}
\label{sec:metrics}



\subsection{Variant Sensitivity Measure (VSM)}

We define
\begin{equation}
\text{VSM} = \frac{\text{Best ASR} - \text{Mean ASR}}{\text{Best ASR}},
\end{equation}
where Best ASR is the maximum ASR across the variant space and Mean ASR is the average ASR across the same space. VSM ranges from 0 (all variants perform identically; the published number is representative) to nearly 1 (only one variant works well; the published number reflects an extreme of the distribution). VSM quantifies how much information is lost when a single number is reported in place of the full variant distribution.



\subsection{Union Coverage (UC)}

We define
\begin{equation}
\text{UC} = \frac{\left| \bigcup_{v \in \mathcal{V}} \mathcal{S}(v) \right|}{|\mathcal{P}|},
\end{equation}
where $\mathcal{V}$ is the variant space, $\mathcal{S}(v)$ is the set of prompts successfully jailbroken by variant $v$, and $\mathcal{P}$ is the full set of evaluated prompts. UC measures the total attack surface accessible to an adversary willing to try multiple variants. The gap between Best ASR and UC quantifies how much the headline number understates the total threat. A UC substantially higher than Best ASR indicates that variants are complementary, that is, they succeed on different subsets of prompts. UC is the metric defenders should use when prioritising patches, because it captures all prompts that any tested variant of the attack family can successfully jailbreak.



\subsection{Reporting Convention}

We argue that any paper proposing or evaluating a parameterised jailbreak attack should report (a) ASR per variant across a clearly enumerated variant space, (b) Best ASR and Mean ASR with VSM computed from them, (c) Union Coverage across the full variant space. This reporting package is sufficient to detect inflation, coverage underestimation, and complementarity. They are inexpensive to compute given that the underlying generations have already been produced, every paper that reports a best-case number must already have generated the variant data.

\section{Case Study 1: PAIR shows template complementarity}
\label{sec:pair}

\subsection{Experimental Setup}
 

\textbf{Attack:} PAIR (Prompt Automatic Iterative Refinement) \citep{chao2025jailbreaking}, a black-box attack in which an attacker LLM iteratively refines prompts based on the target model's responses. \textbf{Dataset:} 100 prompts stratified across the 7 harm categories of HarmBench \citep{mazeika2024harmbench} in proportion to their representation in the full 400-prompt set. \textbf{Target models:} Mistral-7B-Instruct and Qwen3-0.6B, providing two target models with different alignment strengths. \textbf{Attacker model:} Llama-3.1-8B-Instruct~\cite{grattafiori2024llama}. \textbf{Templates tested:} all three system prompt templates released with the original PAIR paper: Roleplay, Logical Appeal, and Authority Endorsement. \textbf{Variant space:} template $\in \{$Roleplay, Logical Appeal, Authority Endorsement$\}$ crossed with conversation rounds $K \in \{1, 2, 3\}$, yielding 9 configurations per target model. We use a single sampling stream ($N=1$) to isolate template and depth effects from best-of-$n$ sampling. \textbf{Judge:} HarmBench classifier (\texttt{cais/HarmBench-Llama-2-13b-cls}).


\subsection{Per-Template ASR and Three-Template Union}



\begin{table}[htbp]
  \caption{PAIR ASR across system prompt templates, conversation depths, and target models. 
  $K$ denotes the number of PAIR refinement rounds. Each template row reports ASR for one system prompt template at each value of $K$; \textit{Best per $K$} reports the strongest single template at that depth; \textit{Mean per $K$} averages across the three templates; and \textit{Union} reports the fraction of prompts jailbroken by at least one of the three templates at that depth.}
  \label{tab:pair-asr}
  \centering
  \small
  \setlength{\tabcolsep}{4pt}
  \begin{tabular}{lcccc|cccc}
    \toprule
    & \multicolumn{4}{c|}{\textbf{Mistral-7B-Instruct}} & \multicolumn{4}{c}{\textbf{Qwen3-0.6B}} \\
    Template & $K=1$ & $K=2$ & $K=3$ & Mean & $K=1$ & $K=2$ & $K=3$ & Mean \\
    \midrule
    Roleplay              & 19\% & 42\% & 53\% & 38.0\% & 28\% & 53\% & 61\% & 47.3\% \\
    Logical Appeal        & 17\% & 31\% & 55\% & 34.3\% & 31\% & 56\% & 64\% & 50.3\% \\
    Authority Endorsement & 24\% & 47\% & 69\% & 46.7\% & 35\% & 61\% & 75\% & 57.0\% \\
    \midrule
    Best per $K$          & \hred{24\%} & \hred{47\%} & \hred{69\%} & --- & \hred{35\%} & \hred{61\%} & \hred{75\%} & --- \\
    Mean per $K$          & 20.0\% & 40.0\% & 59.0\% & --- & 31.3\% & 56.7\% & 66.7\% & --- \\
    \textbf{Union}        & \hgrn{48\%} & \hgrn{75\%} & \hgrn{88\%} & --- & \hgrn{66\%} & \hgrn{88\%} & \hgrn{93\%} & --- \\
    \bottomrule
  \end{tabular}
\end{table}

Table~\ref{tab:pair-asr} shows that ASR varies substantially across PAIR templates, conversation depths, and target models. Increasing the number of refinement rounds consistently improves attack success for all templates on both Mistral-7B-Instruct and Qwen3-0.6B, but no single template captures the full set of vulnerable prompts. Authority Endorsement is the strongest individual template for both targets, reaching 69\% ASR on Mistral-7B and 75\% ASR on Qwen3-0.6B at $K=3$. However, the three-template union is substantially higher than the best individual template at every value of $K$. On Mistral-7B, the union exceeds the best template by 24 percentage points at $K=1$, 28 points at $K=2$, and 19 points at $K=3$. On Qwen3-0.6B, the corresponding gaps are 31, 27, and 18 percentage points. This shows that the templates are complementary: different templates jailbreak different prompts, so reporting per-template ASR alone still does not reveal how much additional prompt coverage is obtained when configurations are combined. In particular, at $K=3$, the best single template suggests 69\% ASR on Mistral-7B and 75\% on Qwen3-0.6B, while trying all three templates reaches 88\% and 93\% coverage, respectively. Moreover, complementarity can occur not only across templates at a fixed $K$, but also across depths: prompts reached by one template at $K=1$ may differ from those reached by another template at $K=3$.


\subsection{Full Variant Space and VSM}

The previous subsection showed that PAIR templates are complementary in terms of prompt coverage. We now ask a different question: how representative is the best reported configuration of the full tested variant space? For each target model, we compute the full-space mean ASR by averaging over all 9 configurations: three templates crossed with three conversation depths. We then compute VSM using the best single configuration and the full-space mean.

\paragraph{Mistral-7B-Instruct.}
For Mistral-7B-Instruct, the best single PAIR configuration is Authority Endorsement at $K=3$, with 69\% ASR. The mean ASR across all 9 tested configurations is 39.7\%. Therefore,
\begin{equation}
  \text{VSM}_{\text{full}} = \frac{69 - 39.7}{69} = \mathbf{0.425}.
  \label{eq:VSM-pair-mistral}
\end{equation}
A full-space VSM of 0.425 means that the best reported configuration is substantially above the average over the tested variant space. In other words, a headline result of 69\% ASR would not represent typical PAIR performance across the configurations we evaluated.

\paragraph{Qwen3-0.6B.}
For Qwen3-0.6B, the best single configuration is again Authority Endorsement at $K=3$, with 75\% ASR. The mean ASR across all 9 tested configurations is 51.6\%. Therefore,
\begin{equation}
  \text{VSM}_{\text{full}} = \frac{75 - 51.6}{75} = \mathbf{0.312}.
  \label{eq:VSM-pair-qwen}
\end{equation}
The lower VSM on Qwen3-0.6B indicates that PAIR performance is more uniform across the tested configurations than on Mistral-7B-Instruct. However, the best configuration still remains noticeably above the full-space mean, showing that even when a model is broadly vulnerable, a single headline ASR can still obscure variation across the variant space.

\subsection{PAIR's Variant Space Is Larger Than We Tested}

We deliberately test PAIR with $N=1$ to isolate the effects of template choice and conversation depth from best-of-$n$ sampling. The full PAIR variant space is larger: it can include the number of parallel streams ($N$), attacker model choice, stopping criteria, and other implementation choices. Our results should therefore be read as a reduced variant-space analysis rather than a complete enumeration of PAIR. Even under this reduced setting, the Union--Best gap remains substantial: at $K=3$, UC exceeds the best single template by 19 percentage points on Mistral-7B and 18 points on Qwen3-0.6B. Both models also show full-space VSM above 0.3, indicating that the best configuration is not fully representative of the tested variant space. These results suggest that template and depth choices alone are enough to create meaningful variation in both ASR and prompt coverage. For attacks with very large parameter spaces such as PAIR, exhaustive enumeration may be impractical. However, this does not weaken the case for distributional reporting. Authors should clearly specify the evaluated variant space and report ASR, VSM, and UC across as much of that space as compute allows.

\section{Case Study 2: Bijection shows extreme variant sensitivity}
\label{sec:bijection}

\subsection{Experimental Setup}

\textbf{Attack:} Bijection learning \citep{huang2024endless}, a black-box attack that teaches the target model a string-to-string encoding via in-context examples and then sends a harmful query in the encoded form. \textbf{Dataset:} The same stratified 100-behavior subset from HarmBench used in Section~\ref{sec:pair}. \textbf{Target models:} Qwen3-0.6B, Mistral-7B-Instruct, and Llama-3.1-8B-Instruct, spanning weak, moderate, and strong alignment. \textbf{Variant space:} dispersion FS $\in \{0, 5, 10, 13, 18, 26\}$ crossed with teaching shots TS $\in \{1, 2, 4, 6, 8, 10\}$, yielding 36 configurations per model. \textbf{Judge:} HarmBench classifier (same as PAIR experiments). \textbf{Total generations:} 3{,}600 per model (100 behaviors $\times$ 36 configurations).

\subsection{Headline Findings}


\begin{table}[htbp]
  \caption{Summary metrics for bijection learning across three target models. FS denotes fixed-size dispersion, i.e., the number of characters left unchanged in the encoded prompt, so the effective dispersion is $26-\text{FS}$. Best ASR, Mean ASR, VSM, and Union Coverage follow the definitions in Section~\ref{sec:metrics}.}
  \label{tab:bij-summary}
  \centering
  \begin{tabular}{lccccc}
    \toprule
    Model & Best ASR & Mean ASR & Worst ASR & VSM & Union Coverage \\
    \midrule
    Qwen3-0.6B   & \hred{52.0\%} & 7.08\%  & 0.0\% & 0.864 & \hgrn{85.0\% (85/100)} \\
    Mistral-7B   & \hred{81.0\%} & 15.94\% & 0.0\% & 0.803 & \hgrn{100.0\% (100/100)} \\
    Llama-3.1-8B & \hred{17.0\%} & 2.69\%  & 0.0\% & 0.842 & \hgrn{43.0\% (43/100)} \\
    \bottomrule
  \end{tabular}
\end{table}

Table~\ref{tab:bij-summary} shows that bijection learning is highly sensitive to variant choice across all three target models. Although the best single-configuration ASR varies widely, from 17\% on Llama-3.1-8B to 81\% on Mistral-7B, VSM remains consistently high at 0.80--0.86. This means that the best reported configuration is far above the mean across the tested variant space for every model. In other words, the headline ASR is not representative of the attack family as a whole.

The Union Coverage column shows the complementary problem: the total prompt set vulnerable to at least one tested configuration is much larger than the best single-configuration ASR. On Qwen3-0.6B, UC reaches 85\% although the best variant reaches 52\%. On Mistral-7B, UC reaches 100\%, meaning that every evaluated prompt is jailbroken by at least one of the 36 tested configurations, even though the best single variant covers only 81\%. Even on Llama-3.1-8B, where the best ASR is only 17\%, UC rises to 43\%. Thus, single-configuration reporting can simultaneously make an attack appear less broadly variable and make a model appear less vulnerable than it is across the tested variant space.

\subsection{Teaching-Shot Sensitivity on Mistral-7B}
\label{sec:bij-mistral-grid}

\begin{table}[htbp]
  \caption{Full ASR grid for bijection learning on Mistral-7B-Instruct. Rows vary dispersion (FS), columns vary teaching shots (TS), and each cell reports ASR for one configuration. The final row and final column report Union Coverage across teaching-shot values and dispersion values, respectively.}
  \label{tab:bij-mistral-grid}
  \centering
  \small
  \begin{tabular}{lccccccc}
    \toprule
    FS \textbackslash{} TS & TS=1 & TS=2 & TS=4 & TS=6 & TS=8 & TS=10 & Union \\
    \midrule
    FS=0  & 17.0\% & 0.0\%  & 0.0\%  & 0.0\%  & 0.0\%  & 0.0\%  & \hgrn{17.0\%} \\
    FS=5  & 28.0\% & 0.0\%  & 0.0\%  & 0.0\%  & 0.0\%  & 0.0\%  & \hgrn{28.0\%} \\
    FS=10 & 24.0\% & 0.0\%  & 0.0\%  & 0.0\%  & 0.0\%  & 0.0\%  & \hgrn{24.0\%} \\
    FS=13 & 15.0\% & 2.0\%  & 2.0\%  & 1.0\%  & 2.0\%  & 2.0\%  & \hgrn{18.0\%} \\
    FS=18 & 23.0\% & 12.0\% & 12.0\% & 15.0\% & 13.0\% & 12.0\% & \hgrn{42.0\%} \\
    FS=26 & 81.0\% & 80.0\% & 51.0\% & 70.0\% & 63.0\% & 49.0\% & \hgrn{95.0\%} \\
    \midrule
    \textbf{Union} & 96.0\% & 84.0\% & 57.0\% & 76.0\% & 67.0\% & 55.0\% & \hgrn{\textbf{100.0\%}} \\
    \bottomrule
  \end{tabular}
\end{table}

Table~\ref{tab:bij-mistral-grid} shows that bijection success on Mistral-7B depends strongly on the interaction between fixed-size dispersion (FS) and teaching-shot count (TS). The best single configuration, FS=26 with TS=1, reaches 81\% ASR, but this does not capture the full variant space. At the same FS value, ASR ranges from 49\% to 81\% depending on TS, showing that teaching-shot count alone can change measured success substantially.

The TS=1 column also reveals failures that would be missed under other teaching-shot settings. Even configurations with small FS values reach 17--28\% ASR at TS=1, while these signals nearly vanish when TS$\geq$2. This indicates that some prompts are vulnerable only under specific parameter interactions. As a result, FS=26 alone covers 95\% of prompts across TS values, but the full union across all 36 configurations reaches 100\%. Thus, no single configuration fully captures the attack surface on Mistral-7B.

\section{Discussion}
\label{sec:discussion}

\subsection{Why Distributional Reporting Helps Defenders}

Single-value ASR is built on an attacker's-best-case worldview: it captures the maximum success an attacker can achieve under one selected configuration. This is one valid threat model, but it is not the only one. A defender deploying a model in the real world cares not only about the strongest single configuration, but also about the total set of prompts that can be jailbroken across the variants an adversary might try.

Our PAIR and bijection results show why this distinction matters. In both attack families, the best single configuration does not cover the full reachable attack surface: other templates, depths, dispersions, or teaching-shot settings expose additional vulnerable prompts. A defender who reasons only from the headline ASR may therefore patch against the reported configuration while missing prompts that remain vulnerable under other variants of the same attack family. VSM captures whether the headline ASR is representative of the tested variant space, while UC captures the residual attack surface exposed beyond any single configuration. Neither can be inferred from a single ASR value.

\subsection{Why Distributional Reporting Helps Attack Researchers}

Researchers proposing new attacks also benefit from distributional reporting. Two attacks reporting ASRs of 80\% and 50\% might appear to have a clear ranking: the first seems stronger than the second. However, if the first has VSM $=0.8$ and the second has VSM $=0.2$, the comparison is less straightforward. The first reflects a best-case extreme, while the second is closer to its mean performance across tested variants. Their headline numbers are therefore not on the same scale. Distributional reporting makes the comparison more interpretable. It also reveals where attack improvements come from: a new attack with higher Best ASR but the same VSM has improved the extreme case, not the mean-case threat. A new attack with lower VSM represents progress in consistency across the variant space.

\subsection{Computational Cost of Distributional Reporting}

A possible concern is that distributional reporting imposes additional compute burden. We agree that exhaustive enumeration may be impractical for attacks with very large parameter spaces such as PAIR or GCG. However, our proposal does not require full enumeration. It requires authors to report the variant space they actually evaluated, along with per-variant ASR, VSM, and UC over that evaluated space. In many cases, authors already run multiple configurations to select or justify a reported setting, so the marginal cost is mainly fuller reporting rather than entirely new experimentation. Even when only a subset of configurations can be tested, reporting that subset is still substantially more informative than reporting one selected point. The marginal cost is a small amount of additional analysis; the benefit is that subsequent work can compare attacks more fairly and defenders can better estimate the reachable attack surface.



\subsection{``Best-case ASR is the right thing to report; it represents what an attacker can actually do.''}

This is the strongest counterargument. Best ASR is operationally meaningful because it captures the maximum observed threat under the strongest tested configuration. We do not propose abandoning it. Our claim is not that Best ASR is incorrectly computed. Best ASR is the maximum ASR over the tested configurations and is therefore the correct summary of the strongest single configuration. The limitation is that this maximum does not capture complementarity across configurations: different variants may succeed on different prompts, so the union of successful prompts across variants can be larger than the success set of the best variant alone.

We therefore propose \textit{complementing} Best ASR with distributional reporting. A single Best ASR value cannot distinguish an attack that works consistently across variants (low VSM, high UC, low Union--Best gap) from an attack that succeeds only after extensive variant search (high VSM), or from an attack whose total prompt coverage is much higher than its Best ASR suggests (high Union--Best gap). These are operationally different threat profiles, and distributional reporting makes the distinction visible.




\subsection{``This is a measurement nuance; researchers can already infer it from existing papers.''}

Some papers do report parameter ablations, such as dispersion curves for bijection learning or $N \times K$ ablations for PAIR. However, ablations within a single paper are usually designed to justify parameter choices, not to support distributional threat reporting. They may show ASR for several variants, but they typically do not report prompt-level overlap across variants, which means Union Coverage cannot be recovered from the published numbers alone. They also do not provide a common basis for cross-paper comparison. When one paper reports best-case ASR for attack X and another reports best-case ASR for attack Y, the two numbers may not be comparable if the underlying attacks have different variant sensitivity. VSM and UC make this hidden variation explicit. Without such metrics, cross-paper comparisons remain unreliable in a way that no individual paper's internal ablation can fully resolve. A community-wide reporting convention is therefore needed.

\section{Limitations}
\label{sec:limitations}

Our empirical evidence covers two attack families evaluated on open-source target models, using a 100-prompt stratified subset of HarmBench. We expect similar distributional effects in other parameterised attacks, such as AutoDAN, GCG, TAP, and CodeAttack, but broader cross-attack validation remains future work. Our PAIR study uses $N=1$ to isolate template and depth effects from best-of-$n$ sampling; varying $N$, attacker model choice and stopping criteria may reveal additional sources of variant sensitivity. We do not evaluate closed-source frontier models because exhaustive variant-level evaluation would require substantial API expenditure, especially when computing Union Coverage across many configurations. Despite these limitations, the consistency of our findings, Union--Best gaps across PAIR on two target models, and VSM between 0.80 and 0.86 across three open-source models in bijection learning, suggests that the position is likely to generalise beyond our specific empirical scope.



\section{Conclusion}
\label{sec:conclusion}

Single-configuration ASR reporting is convenient but insufficient. It captures one configuration of a parameterised attack and discards information about whether that configuration is representative of the tested variant space and how much of the attack surface it captures. We have provided empirical evidence that this matters: across two attack families and three target models, the gap between best-case ASR and Union Coverage ranges from 18 to 33 percentage points. We propose two metrics, VSM and UC, that complement ASR and require only fuller reporting of the configurations already evaluated. Reporting them would help defenders prioritise vulnerable prompts, help researchers compare attacks more fairly, and move the field from headline-driven evaluation toward a more complete characterisation of jailbreak threat. We therefore urge the community to treat distributional ASR reporting as a minimum standard for evaluating parameterised jailbreak attacks.



\bibliographystyle{plainnat}
\bibliography{references}

@inproceedings{chao2025jailbreaking,
  title={Jailbreaking black box large language models in twenty queries},
  author={Chao, Patrick and Robey, Alexander and Dobriban, Edgar and Hassani, Hamed and Pappas, George J and Wong, Eric},
  booktitle={2025 IEEE Conference on Secure and Trustworthy Machine Learning (SaTML)},
  pages={23--42},
  year={2025},
  organization={IEEE}
}

@article{huang2024endless,
  title={Endless jailbreaks with bijection learning},
  author={Huang, Brian RY and Li, Maximilian and Tang, Leonard},
  journal={arXiv preprint arXiv:2410.01294},
  year={2024}
}

@article{zou2023universal,
  title={Universal and transferable adversarial attacks on aligned language models},
  author={Zou, Andy and Wang, Zifan and Carlini, Nicholas and Nasr, Milad and Kolter, J Zico and Fredrikson, Matt},
  journal={arXiv preprint arXiv:2307.15043},
  year={2023}
}

@article{liu2023autodan,
  title={Autodan: Generating stealthy jailbreak prompts on aligned large language models},
  author={Liu, Xiaogeng and Xu, Nan and Chen, Muhao and Xiao, Chaowei},
  journal={arXiv preprint arXiv:2310.04451},
  year={2023}
}

@article{mazeika2024harmbench,
  title={Harmbench: A standardized evaluation framework for automated red teaming and robust refusal},
  author={Mazeika, Mantas and Phan, Long and Yin, Xuwang and Zou, Andy and Wang, Zifan and Mu, Norman and Sakhaee, Elham and Li, Nathaniel and Basart, Steven and Li, Bo and others},
  journal={arXiv preprint arXiv:2402.04249},
  year={2024}
}

@article{mehrotra2024tree,
  title={Tree of attacks: Jailbreaking black-box llms automatically},
  author={Mehrotra, Anay and Zampetakis, Manolis and Kassianik, Paul and Nelson, Blaine and Anderson, Hyrum and Singer, Yaron and Karbasi, Amin},
  journal={Advances in Neural Information Processing Systems},
  volume={37},
  pages={61065--61105},
  year={2024}
}

@inproceedings{ren2024codeattack,
  title={Codeattack: Revealing safety generalization challenges of large language models via code completion},
  author={Ren, Qibing and Gao, Chang and Shao, Jing and Yan, Junchi and Tan, Xin and Lam, Wai and Ma, Lizhuang},
  booktitle={Findings of the Association for Computational Linguistics: ACL 2024},
  pages={11437--11452},
  year={2024}
}

@article{souly2024strongreject,
  title={A strongreject for empty jailbreaks},
  author={Souly, Alexandra and Lu, Qingyuan and Bowen, Dillon and Trinh, Tu and Hsieh, Elvis and Pandey, Sana and Abbeel, Pieter and Svegliato, Justin and Emmons, Scott and Watkins, Olivia and others},
  journal={Advances in Neural Information Processing Systems},
  volume={37},
  pages={125416--125440},
  year={2024}
}

@article{xu2024bag,
  title={Bag of tricks: Benchmarking of jailbreak attacks on llms},
  author={Xu, Zhao and Liu, Fan and Liu, Hao},
  journal={Advances in Neural Information Processing Systems},
  volume={37},
  pages={32219--32250},
  year={2024}
}

@article{curl2026beyond,
  title={Beyond Attack Success Rate: A Multi-Metric Evaluation of Adversarial Transferability in Medical Imaging Models},
  author={Curl, Emily and Ampomah, Kofi and Erfan, Md and Dibbo, Sayanton},
  journal={arXiv preprint arXiv:2604.16532},
  year={2026}
}

@article{yang2025qwen3,
  title={Qwen3 technical report},
  author={Yang, An and Li, Anfeng and Yang, Baosong and Zhang, Beichen and Hui, Binyuan and Zheng, Bo and Yu, Bowen and Gao, Chang and Huang, Chengen and Lv, Chenxu and others},
  journal={arXiv preprint arXiv:2505.09388},
  year={2025}
}

@article{grattafiori2024llama,
  title={The llama 3 herd of models},
  author={Grattafiori, Aaron and Dubey, Abhimanyu and Jauhri, Abhinav and Pandey, Abhinav and Kadian, Abhishek and Al-Dahle, Ahmad and Letman, Aiesha and Mathur, Akhil and Schelten, Alan and Vaughan, Alex and others},
  journal={arXiv preprint arXiv:2407.21783},
  year={2024}
}

@article{mistral7b,
  title     = {Mistral {7B}},
    author = {Jiang, Albert Q. and Sablayrolles, Alexandre and Mensch, Arthur and
          Bamford, Chris and Chaplot, Devendra Singh and {de las Casas}, Diego and
          Bressand, Florian and Lengyel, Gianna and Lample, Guillaume and
          Saulnier, Lucile and {Renard Lavaud}, L{\'e}lio and Lachaux, Marie-Anne and
          Stock, Pierre and {Le Scao}, Teven and Lavril, Thibaut and Wang, Thomas and
          Lacroix, Timoth{\'e}e and {El Sayed}, William},
  year      = {2023},
    url       = {https://arxiv.org/abs/2310.06825}
}



\appendix

\section{Appendix}
\subsection{ASR Grids for Tested Bijection Configurations}
\label{app:grids}

This appendix reports the ASR grids for the tested bijection-learning configurations on Qwen3-0.6B and Llama-3.1-8B-Instruct. Rows correspond to fixed-size dispersion (FS), columns correspond to the number of teaching shots (TS), and each cell reports the ASR for one evaluated configuration. The final column reports Union Coverage across teaching-shot values for a fixed FS, while the final row reports Union Coverage across FS values for a fixed TS. The bottom-right cell gives the Union Coverage across all 36 tested configurations.

\paragraph{Qwen3-0.6B.}

\begin{table}[htbp]
  \caption{ASR grid for tested bijection configurations on Qwen3-0.6B.}
  \centering
  \small
  \begin{tabular}{lccccccc}
    \toprule
    FS \textbackslash{} TS & TS=1 & TS=2 & TS=4 & TS=6 & TS=8 & TS=10 & Union \\
    \midrule
    FS=0  & 0.0\%  & 0.0\%  & 0.0\%  & 0.0\%  & 0.0\%  & 0.0\%  & \hgrn{0.0\%} \\
    FS=5  & 0.0\%  & 0.0\%  & 0.0\%  & 0.0\%  & 0.0\%  & 0.0\%  & \hgrn{0.0\%} \\
    FS=10 & 0.0\%  & 0.0\%  & 0.0\%  & 0.0\%  & 0.0\%  & 0.0\%  & \hgrn{0.0\%} \\
    FS=13 & 0.0\%  & 0.0\%  & 0.0\%  & 0.0\%  & 0.0\%  & 0.0\%  & \hgrn{0.0\%} \\
    FS=18 & 4.0\%  & 2.0\%  & 4.0\%  & 2.0\%  & 3.0\%  & 3.0\%  & \hgrn{12.0\%} \\
    FS=26 & 37.0\% & 33.0\% & 37.0\% & 52.0\% & 39.0\% & 39.0\% & \hgrn{83.0\%} \\
    \midrule
    \textbf{Union} & 40.0\% & 33.0\% & 39.0\% & 52.0\% & 40.0\% & 39.0\% & \hgrn{\textbf{85.0\%}} \\
    \bottomrule
  \end{tabular}
\end{table}

For Qwen3-0.6B, successful jailbreaks are concentrated at larger FS values, especially FS=26. The best single configuration reaches 52.0\% ASR, but the full union across all configurations reaches 85.0\%. This shows that multiple configurations contribute non-overlapping successes, even though many low-FS configurations individually achieve 0.0\% ASR.

\paragraph{Llama-3.1-8B-Instruct.}

\begin{table}[htbp]
  \caption{ASR grid for tested bijection configurations on Llama-3.1-8B-Instruct.}
  \centering
  \small
  \begin{tabular}{lccccccc}
    \toprule
    FS \textbackslash{} TS & TS=1 & TS=2 & TS=4 & TS=6 & TS=8 & TS=10 & Union \\
    \midrule
    FS=0  & 0.0\% & 0.0\%  & 0.0\%  & 0.0\%  & 0.0\% & 0.0\% & \hgrn{0.0\%} \\
    FS=5  & 0.0\% & 0.0\%  & 0.0\%  & 0.0\%  & 0.0\% & 0.0\% & \hgrn{0.0\%} \\
    FS=10 & 0.0\% & 0.0\%  & 0.0\%  & 0.0\%  & 0.0\% & 0.0\% & \hgrn{0.0\%} \\
    FS=13 & 0.0\% & 0.0\%  & 0.0\%  & 0.0\%  & 0.0\% & 0.0\% & \hgrn{0.0\%} \\
    FS=18 & 6.0\% & 5.0\%  & 4.0\%  & 7.0\%  & 8.0\% & 5.0\% & \hgrn{16.0\%} \\
    FS=26 & 1.0\% & 17.0\% & 17.0\% & 13.0\% & 7.0\% & 7.0\% & \hgrn{31.0\%} \\
    \midrule
    \textbf{Union} & 6.0\% & 20.0\% & 19.0\% & 18.0\% & 14.0\% & 12.0\% & \hgrn{\textbf{43.0\%}} \\
    \bottomrule
  \end{tabular}
\end{table}

For Llama-3.1-8B-Instruct, individual ASR values are lower than for Qwen3-0.6B, with the best single configuration reaching only 17.0\%. Nevertheless, the full union reaches 43.0\%, more than double the best single-configuration ASR. This again illustrates the central point of the paper: a low headline ASR can still conceal a larger set of prompts that are vulnerable under other tested configurations.

\end{document}